\documentclass{jfm}

\ifCUPmtlplainloaded \else
  \checkfont{eurm10}
  \iffontfound
    \IfFileExists{upmath.sty}
      {\typeout{^^JFound AMS Euler Roman fonts on the system,
                   using the 'upmath' package.^^J}%
       \usepackage{upmath}}
      {\typeout{^^JFound AMS Euler Roman fonts on the system, but you
                   dont seem to have the}%
       \typeout{'upmath' package installed. JFM.cls can take advantage
                 of these fonts,^^Jif you use 'upmath' package.^^J}%
      }
  \else
  \fi
\fi

\ifCUPmtlplainloaded \else
  \checkfont{msam10}
  \iffontfound
    \IfFileExists{amssymb.sty}
      {\typeout{^^JFound AMS Symbol fonts on the system, using the
                'amssymb' package.^^J}%
       \usepackage{amssymb}%
       \let\le=\leqslant  \let\leq=\leqslant
         
      }{}
  \fi
\fi

\ifCUPmtlplainloaded \else
  \IfFileExists{amsbsy.sty}
    {\typeout{^^JFound the 'amsbsy' package on the system, using it.^^J}%
     \usepackage{amsbsy}}
    {}
\fi

\usepackage[dvips]{graphicx}

\title[Anomalous scaling of low-order structure functions] {Anomalous scaling of low-order
structure functions of turbulent velocity}

\author[S.Y. Chen, B. Dhruva, S. Kurien, K. R. Sreenivasan and M.A. Taylor]
{S.\ns Y.\ns  C\ls H\ls E\ls N$^1$,\ns B.\ns D\ls H\ls R\ls U\ls
V\ls A$^2$,\ns S.\ns K\ls U\ls R\ls I\ls E\ls N$^3$,\break K.\ns
R.\ns S\ls R\ls E\ls E\ls N\ls I\ls V\ls A\ls S\ls A\ls N$^4$\ns
\and M.\ns A.\ns T\ls A\ls Y\ls L\ls O\ls R$^5$}
\affiliation{$^1$Department of Mechanical Engineering, Johns
Hopkins
University, Baltimore, MD 21218\\[\affilskip] $^2$Schlumberger Cambridge
Research, Madingley Road, Cambridge CB3 0EL\\[\affilskip] $^3$Center for Nonlinear Studies (CNLS)
and Mathematical Modeling \& Analysis Group (T7), Los Alamos
National Laboratory,
Los Alamos, NM 87505\\[\affilskip] $^4$International Centre for Theoretical
Physics, Strada Costiera 11, 34014 Trieste, Italy \\[\affilskip] $^5$Computer and
Computational Sciences (CCS-3), Los Alamos National Laboratory,\\
Los Alamos, NM 87505}

\pubyear{} \volume{} \pagerange{}
\date{\today}
\begin{document}
\maketitle

\begin{abstract}
It is now believed that the scaling exponents of moments of
velocity increments are anomalous, or that the departures from
Kolmogorov's (1941) self-similar scaling increase nonlinearly with
the increasing order of the moment. This appears to be true
whether one considers velocity increments themselves or their
absolute values. However, moments of order lower than 2 of the
absolute values of velocity increments have not been investigated
thoroughly for anomaly. Here, we discuss the importance of the
scaling of non-integer moments of order between +2 and $-1$, and
obtain them from direct numerical simulations at moderate Reynolds
numbers (Taylor microscale Reynolds numbers $R_\lambda \le$ 450)
and experimental data at high Reynolds numbers ($R_\lambda
\approx$ 10,000). The relative difference between the measured
exponents and Kolmogorov's prediction increases as the moment
order decreases towards $-1$, thus showing that the anomaly that
is manifest in high-order moments is present in low-order moments
as well. This conclusion provides a motivation for seeking a
theory of anomalous scaling as the order of the moment vanishes.
Such a theory does not have to consider rare events---which may be
affected by non-universal features such as shear---and so may be
regarded as advantageous to consider and develop.
\end{abstract}

\section{Introduction}
\setlength{\parskip}{0cm}
The moments of velocity differences over spatial scales of size
$r$, the so-called structure functions, provide useful measures of
the statistical description of fluid turbulence (Kolmogorov
1941a,b). In particular, the longitudinal structure functions
defined as
\begin{equation}
S_n({\bf r}) = \Big \langle [({\bf u}({\bf x}+{\bf r}) - {\bf
u}({\bf x}))\cdot {\bf \hat r}]^n \Big \rangle \label{Sn}
\end{equation}
have been studied extensively. Here, $\bf u(\bf x)$ is the
velocity vector at position $\bf x$, and $\bf \hat r$ is the unit
vector along the separation vector $\bf r$. The special interest
in structure functions comes in part from an exact result, known
as the 4/5-ths law,
\begin{equation}
S_3({\bf r}) = - \frac{4}{5} \epsilon r, \label{45th}
\end{equation}
valid in the inertial range of scales ($\eta \ll r \ll L$ where
$\eta$ is the Kolmogorov scale characterizing the dissipative
scale of motion and $L$ is a suitable large scale of turbulence).
In part, the interest is spurred by the operational ease with
which longitudinal structure functions can be obtained from
experimental data if one makes the so-called Taylor's hypothesis
(Taylor 1935). The major impetus for measurements, however, is the
scaling result of Kolmogorov---K41 for brevity---that, for high
Reynolds numbers, the structure functions follow the relation
$S_n({\bf r}) \sim r^{\zeta_n}$ where the scaling exponent
$\zeta_n = {n/3}$. As a result of considerable work (see, for
example, Anselmet {\it et al.} 1984, Maurer {\it et al.} 1994,
Arneodo {\it et al.} 1996, Sreenivasan \& Antonia 1997), it now
appears nearly certain that the scaling exponents deviate from
$n/3$ increasingly and nonlinearly as $n$ increases. This is the
anomalous scaling. While some issues remain to be explained
satisfactorily (see, for example, Sreenivasan \& Dhruva 1998), it
appears that anomalous scaling is a genuine result worthy of a
serious theoretical effort. Consequently, considerable thrusts of
research have occurred in this direction (e.g., L'vov \& Procaccia
1996).

One obvious concern about high-order moments is that, since they
sample the tails of the probability distribution function (PDF) of
velocity increments---and since some of the associated rare events
may be related to well-defined flow structures in real space---it
is not entirely certain that the results for high-order moments
are universal because the mean shear and other non-universal
features driving the flow could influence the rare structures of
turbulence. In contrast, low-order moments are less susceptible to
such non-universal aspects, because they are determined nearly
entirely by the core of the PDF and hence sampled frequently.
Thus, it is reasonable to regard anomalous scaling---and its
universality---as more conclusively established if low-order
moments also display anomaly. This is the main subject of this
paper.

The lowest non-trivial structure function that has been studied
extensively is the second-order, whose scaling exponent has been
shown to be $\approx 0.7$. Though this is measurably different
from the predicted value of $2/3$, the difference is too small to
be conclusive on its own. It would thus be useful to examine
scaling exponents for moments of still lower orders. Since moments
of order $-1$ and below do not exist for velocity increments
(\cite{cast90}), the range of our current interest is limited to
$-1 < n \le 2$, where $n$ is necessarily fractional. With
decreasing $n$ in this range, if the deviation from $n/3$
decreases, we shall at least know that K41 will be exact in the
limit of low-order moments, and regard it as a possible reference
point for a theory. If, on the other hand, these deviations remain
to be non-trivial as $n$ vanishes, it may well be that the
understanding of the anomaly can be sought more fruitfully in
terms of low-order moments, for the simple reason that such a
theory can justifiably ignore rare events. Some preliminary
measurements were published in Sreenivasan {\it et al.} (1996) and
Cao {\it et al.} (1996) but the present paper is a more complete
account of the data. Perhaps more importantly, the preliminary
numbers did not take account of the possible effects of residual
anisotropy in both experiments and simulations, an issue whose
importance has been recently highlighted (Biferale \& Procaccia
2004). We take account of this feature using a recently developed
angle-averaging technique (Taylor {\it et al.} 2003).

The rest of the paper is organized as follows. In Sec.\ 2, we
describe the experimental and numerical data used for the present
analysis. This is followed in Sec.\ 3 by the calculation of the
fractional structure functions and their scaling exponents from
the datasets described in Sec.\ 2. The exponents from all sources
of data agree well with each other and deviate measurably from the
K41 prediction even as $n \rightarrow 0$. Section 4 contains a
brief discussion of the significance of the results.

\section{Experimental and numerical data}

\subsection{High-Reynolds number atmospheric boundary layer measurements}

Hotwire measurements were made in the atmospheric surface layer at
a height of 35 m above the ground using a standard meteorological
tower at Brookhaven National Laboratory. The tower itself
presented very little obstacle to the wind because of its low
solidity. The dataset analyzed here is part of a more
comprehensive batch obtained at the tower. The hotwire, 0.7 mm in
length and 0.5 $\mu$m in diameter, was placed facing the wind,
about two meters away from the tower. (For monitoring the wind
direction, the tower was equipped with a vane anemometer placed
two meters away from the measurement station.) The calibration was
performed in situ using a TSI calibrator and checked later in a
windtunnel. The signals were low-pass filtered at 5 kHz and
sampled at 10 kHz. The anemometer and signal conditioners were
placed nearby at the height of measurement, and the conditioned
signal was transmitted to the ground and digitized using a 12-bit
A/D converter. Typical data records contained between 10 and 40
million samples, during which time the wind direction and its mean
speed were deemed acceptably constant. More details are given in
\cite{bd00}, but the essential features for this particular set of
data are listed in table \ref{expt_params}. The wind conditions
were somewhat unstable.

\begin{table}
\begin{center}
\begin{tabular}{cccccc}
U & u$^{\prime}$& $\epsilon$ & $\eta$ & $\lambda$ & $R_\lambda$\\
7.6 ms$^{-1}$& 1.36 ms$^{-1}$ & 0.032 m$^2$s$^{-3}$ & 0.57 mm & 11.4 mm &10,340\\
\end{tabular} \caption{Some relevant parameters for the
atmospheric data. Here, $U$ is the mean speed, u$^{\prime}$ is the
root-mean-square velocity, $\epsilon$ is the mean rate of energy
dissipation, $\eta$ and $\lambda$ are the Kolmogorov and Taylor
microscales, respectively, and $R_\lambda \equiv
u^{\prime}\lambda/\nu$, $\nu$ being the kinematic viscosity of air
at the measurement temperature.} \label{expt_params}
\end{center}
\end{table}

\subsection{Direct numerical simulations (DNS) of Navier-Stokes equations with
forcing}\label{chen-data}

The Navier-Stokes equations were solved numerically for periodic
boundary conditions. A pseudospectral code was used using a
second-order time-integration scheme (Cao {\it et al.} 1996).
Simulations were carried out with a resolution of 512$^3$ grid
points on the CM-5 at Los Alamos national Laboratory and SP
machines at IBM Watson Research Center. To obtain a statistically
steady state, a forcing is applied to the first wave-number shells
$0.5 < k < 1.5$ so that at each time step the total energy of that
shell is constant. The maximum $R_\lambda$ was about $250$. Time
integration up to 60 large-eddy turn-over times was performed.

Even though the $512^3$ data show well-developed scaling range in
an ESS plot, the Reynolds number is not large enough to produce
unambiguous scaling in direct log-log plots of moments versus the
scale $r$. It was also found that, even though the turbulence is
nominally isotropic, there are some measurable (though small)
differences between the scaling of longitudinal and transverse
structure functions, suggesting a possible presence of residual
anisotropy in the inertial range, arising from possible anisotropy
in forcing. For these two reasons, it seemed worth considering new
data from a higher Reynolds number simulations, for which the
anisotropy effects, to the extent that they were present at all,
were accounted for in a rational manner. We therefore used the
velocity data from a simulation of the Navier-Stokes equation in a
periodic domain of size $1024^3$. The forcing scheme is similar to
that described for the $512^3$ simulation and is explained in
detail in Taylor {\it et al.} (2003). The steady state was
achieved in about 1.5 large-eddy turnover times and the simulation
ran for a total of 2.5 large-eddy turnover times. The statistical
analysis was performed over 10 frames in the final eddy turnover
time. The steady state value of $R_\lambda$ was 450. Other
parameters of this simulation are given in table \ref{sim_parms}.
\begin{table}
\begin{center}
\begin{tabular}{ccccc}
N & $\nu$ & $\epsilon$ & $\delta x \over \eta$ & $R_\lambda$\\
[3pt] 1024 & $3.5\times 10^{-5}$ &  1.75 & 0.75 & 450\\
\end{tabular} \caption{Some relevant parameters for the $1024^3$
DNS. The fourth column provides the resolution of the simulation
in terms of the Kolmogorov scale.} \label{sim_parms}
\end{center}
\end{table}

\section{Results}
\begin{figure}
\centering
\includegraphics[scale = 0.4]{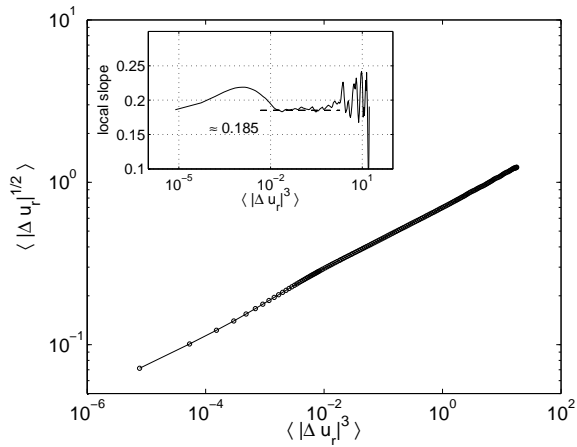}
\caption{ESS calculation of scaling exponent of order 0.5 from
  experimental data. \label{ESS_exp}}
\end{figure}
Since we are interested in real-valued structure functions, we can
consider, when the moment order is either fractional or negative, only
the moments of absolute values of velocity differences defined as
\begin{equation}
S_{|n|}({\bf r}) = \Big \langle \Big|({\bf u}({\bf
    x}+{\bf r}) - {\bf u}({\bf x}))\cdot {\bf \hat r}\Big |^{n} \Big
  \rangle. \label{Sn_a}
\end{equation}
In experimental measurements, because the Reynolds number is quite
high (see table 1), and the scaling ranges reasonably clear, we
have the luxury of estimating the scaling exponent directly from
log-log plots of $S_{|n|}({\bf r})$ versus $r$ (see Dhruva 2000).

We also performed the calculation using extended self-similarity
or ESS (Benzi {\it et al.} 1993). In ESS, a structure function
$S_n({\bf r})$ of interest is plotted against another structure
function $S_m({\bf r})$ and the relative scaling exponent
$\zeta_n/\zeta_m$ is obtained. In particular, if the exponent of
$S_m({\bf r})$ is known {\it a priori}, as from theory for the
third-order structure function (see equation \ref{45th}), then the
exponent of $S_n({\bf r})$ may be inferred. It is known that ESS
improves the scaling range significantly. In the present case we
use $S_m({\bf r}) = S_{|3|}({\bf r})$, the absolute third-order
structure function and assume its scaling exponent is 1. It is
unclear if this assumption is valid since the K41 theory does not
apply to moments of absolute differences; indeed, it appears that
the scaling exponents of $S_3$ and $S_{|3|}$ are slightly
different (Sreenivasan {\it et al.} 2000). However, we found that
the ESS exponents (see table \ref{exponents}) were within the
uncertainty of those obtained directly by \cite{bd00}. An example
of ESS plot for moment-order of $0.5$ is shown in figure
\ref{ESS_exp}, along with the local slope (see inset).

For the numerical simulation data with resolution of 512$^3$, the
exponents were obtained only by ESS because the scaling region was
small in direct log-log plots against the scale $r$. These
exponents are listed in table \ref{exponents}.
\begin{table}
\begin{center}
\begin{tabular}{ccccccc}
order & measured  & relative & DNS exponents & relative & DNS
exponents & relative \\
of moments& exponents& difference& ($512^3$) & difference&
($1024^3$)& difference\\
 -0.80& -0.317& 0.189 &-0.313 & 0.174  & - & - \\
 -0.60&  -    &  -     &  -      &   -     & -0.238$\pm$0.002 & 0.188\\
 -0.40&  -    &  -     &  -      &   -     & -0.158$\pm$0.001 & 0.181\\
 -0.20&-0.078& 0.170 &-0.077& 0.155  & -0.078$\pm$0.001 & 0.171\\
 0.10&0.039& 0.170 &0.036& 0.080&0.039$\pm$0.001& 0.155 \\
 0.20&0.076& 0.140 &0.073& 0.095&0.077$\pm$0.001& 0.152 \\
 0.30&0.113& 0.130 &0.112& 0.120&0.115$\pm$0.001& 0.147 \\
 0.40&0.150& 0.125 &0.150& 0.125&0.152$\pm$0.001& 0.143 \\
 0.50&0.187$\pm 0.003$& 0.140 &0.187& 0.122&0.190$\pm$0.001& 0.138 \\
 0.60&0.221& 0.105 &0.223& 0.115&0.226$\pm$0.001& 0.133 \\
 0.70&0.265& 0.136 &0.260& 0.114&0.263$\pm$0.001& 0.128 \\
 0.80&0.292& 0.095 &0.296& 0.110&0.300$\pm$0.001& 0.123 \\
 0.90&0.333& 0.110 &0.332& 0.107&0.340$\pm$0.001& 0.119 \\
 1   &0.372& 0.116 &0.366& 0.098& 0.370$\pm$0.006& 0.111\\
 1.25&0.458& 0.099 &0.452& 0.085& 0.459$\pm$0.006& 0.101 \\
 1.50&0.542& 0.084 &0.536& 0.072& 0.545$\pm$0.006& 0.091 \\
 1.75&0.628& 0.077 &0.619& 0.061& 0.630$\pm$0.006 &0.079 \\
 2.00&0.704$\pm0.003$& 0.061 &0.699& 0.049& 0.712$\pm$0.006& 0.064 \\
\end{tabular} \caption{Scaling exponents from ESS compared with
those from for isotropic turbulence from two sets DNS data. Error
bars are given for the experimental data for two exponents. Those
for the $512^3$ data are given in Cao {\it et al.} 1996.}
\label{exponents}
\end{center}
\end{table}

\subsection{Effects of finite Reynolds number and anisotropy}
The finite Reynolds number of turbulence in numerical work
shortens the inertial range over which scaling exponents may be
discerned with clarity. This remains a constraint because the
applicable theory concerns the limit of Re $\rightarrow \infty$. A
constraint in experimental data is that some large-scale
anisotropy might be present in the range that appears to scale.
The effect could be present even in high-Reynolds-number flows
because the effects of anisotropic forcing penetrate the scaling
range in a subtle but systematic way; see, for example,
\cite{ALP99,KS01}. In numerical simulations, the statistics are
usually calculated with the separation direction ${\bf {\hat r}}$
oriented parallel to a box-side. If there is residual anisotropy
(angular-dependence) in the small-scales, this procedure will bias
the results. Since we are concerned here with delicate results,
this uncertainty has to be eliminated satisfactorily.

\subsection{Recovering isotropic statistics by angle-averaging}
We make use of two recent developments to properly eliminate the
effect of anisotropy in the inertial range of the $1024^3$ data.
First, we now know that isotropic and anisotropic contributions
can be isolated systematically by projecting structure function of
a given order over a particular basis function in its SO(3) group
decomposition (e.g., Arad {\it et al.} 1998, 1999, Kurien {\it et
al.} 2000, Biferale \& Procaccia 2004). This is a useful step to
take even in nominally isotropic turbulence because the effect of
forcing might persist in the nominal scaling range. Second, Taylor
{\it et al.} (2003) developed a method by which the third-order
longitudinal structure function was computed in $many$ directions
of the flow for a given $r$ and the results averaged. The
angle-averaged value of the structure function achieved the
Kolmogorov 4/5 prediction to a remarkable degree, thus providing a
convincing scheme for extracting the isotropic component of a
mildly anisotropic flow. The two procedures achieve the same goal:
the angle-averaging procedure is in effect a projection on to the
isotropic component of the SO(3) rotation group. Since this method
is independent of the order of the structure function, we use it
as described below for fractional statistics.

The angle-averaged isotropic structure function in the
computational domain $D$ is given by
\begin{equation}
S_{|n|}(r) = \frac{1}{\Delta t}\int_{t_0}^{{t_0} +\Delta t} dt\int
\frac{d\Omega_r}{4\pi} \int_D \frac{d{\bf x}}{L^3}~\Big|({\bf u}({\bf
x}+{\bf r}) - {\bf u}({\bf x}))\cdot {\bf \hat r}\Big|^n,
\label{Sn_angave}
\end{equation}
where the usual average for a particular direction of the unit
separation vector ${\bf {\hat r}}$ is followed by a spherical
average of all possible orientations of ${\bf {\hat r}}$ over the
solid angle $\Omega_r$. The long-time average is taken in the
steady state. In units of the large-eddy turnover time, $t_0$ is
1.5 from the start of the simulation and $\Delta t$ is unity.
\begin{figure}
\centering
\includegraphics[scale = 0.4]{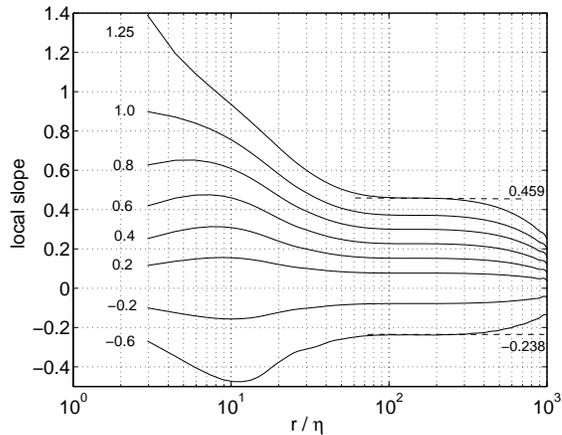}
\caption{The scaling exponents, equal to the local slopes
  $\zeta_{|n|} = {\mbox d}\log (S_{|n|}(r))/{\mbox d}\log(r)$, as a function of $r$,
  for various values of $-1 < n < 2$ for the $1024^3$ DNS data.
  Each curve is labelled on the left by the order of the structure
  function. The scaling exponents deduced in this
  way are given on the right for two representative orders, $n=1.25$
  and $n=-0.6$.} \label{fract_localslp}
\end{figure}

In numerical simulations, since we have the full three-dimensional
velocity field, we can in principle integrate over the sphere and
project out the isotropic part of $S_{|n|}$.  Taylor {\it et al.}
showed that the full spherical average may be approximated to
arbitrary precision by first computing the structure function over
sufficiently many different directions in the flow, interpolating
each of these functions by a simple single-variable cubic spline,
and then averaging the interpolated values over all directions.
This angle-averaging of the structure functions is much faster to
implement than interpolating the three-dimensional,
three-component velocity data over spherical shells in order to
perform the integration. It is also more accurate in the small
scales than other spherical averaging schemes.
\begin{figure}
\centering
\includegraphics[scale = 0.4]{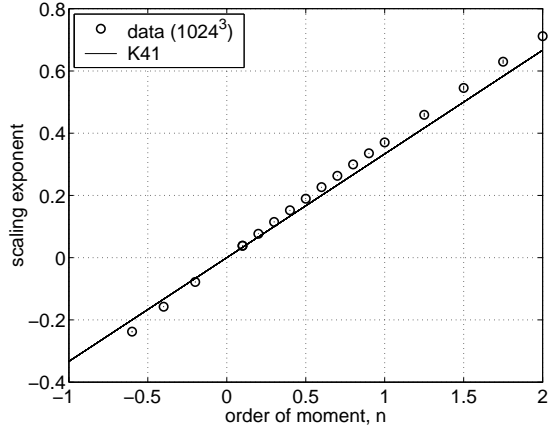}
\caption{The scaling exponents calculated from the $1024^3$ DNS
($\circ$); standard deviations ($\approx \pm$0.001) are smaller
than the size of the circles. Full line: K41 exponents,
extrapolated via self-similarity arguments to low-order
statistics. Anomalous scaling is evident. \label{fract_exps}}
\end{figure}

\begin{figure}
\centering
%\begin{minipage}[t]{6.5cm}
\includegraphics[scale = 0.4]{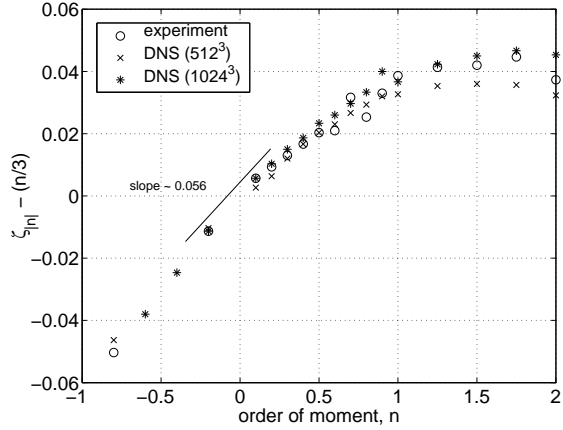}
\caption{The departure from K41 for all three data sets:
experiment ($\circ$), $512^3$ DNS ($\times$) and $1024^3$ DNS
($\star$). The difference goes to zero approximately linearly with
a slope of $\sim 0.056$. This indicates that the departure from
K41 persists all the way to $n=0$. See figure~\ref{relative_exps}
for the relative departure from K41. \label{d-fract_exps}}
\end{figure}
%\end{minipage}
%\hfill
%\begin{minipage}[t]{6.5cm}
\begin{figure}
\centering
\includegraphics[scale = 0.4]{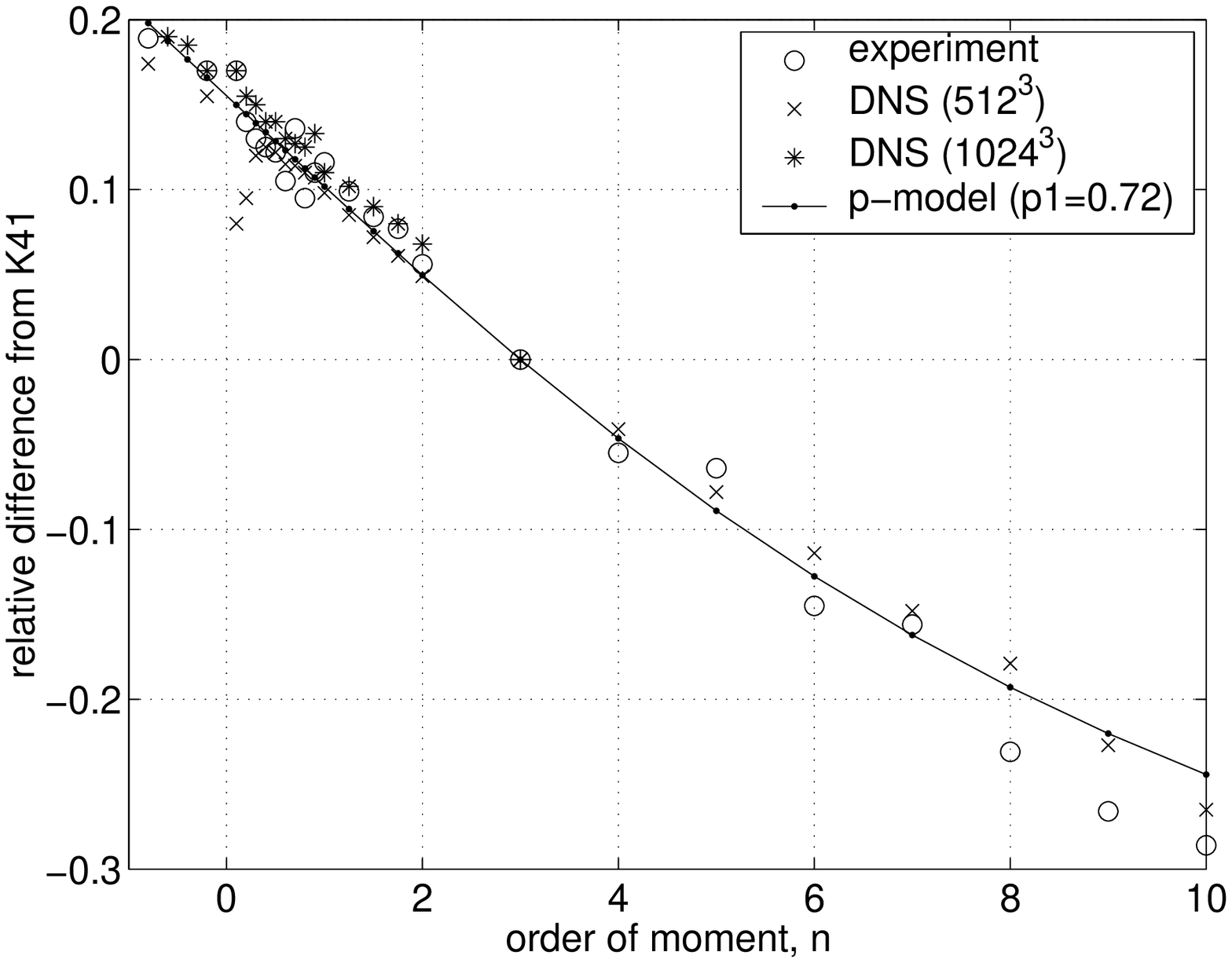}
\caption{The relative difference $(\zeta_{|n|} - n/3)/(n/3)$ for
  the various $-1 < n \leq 10$ as calculated from the experiments
  ($\circ$), DNS ($512^3$) ($\times$) and DNS ($1024^3$) ($\star$).
  This relative difference smoothly goes through $n=0$ without any
  special feature, showing that while the measured exponents also go to
  zero at $n = 0$, they do so at a different rate than the K41 exponents (see
  figures \ref{fract_exps} and \ref{d-fract_exps}). That is, there
  is a finite departure from K41 of $\sim 16\%$ in the limit $n
  \rightarrow 0$.  There is also no special behavior as $n\rightarrow
  -1$ at which point the mathematically defined scaling exponent and
  hence the relative difference from K41, diverges.  The exponents for
  $n>3$ for the experiments and the $512^3$ simulation are taken from
  the values tabulated in \cite{bd00}.
\label{relative_exps}}
%\end{minipage}
%\hfill
\end{figure}

The angle-averaging procedure was performed for all orders. For
very low-order moments $|n| < 1$ we observed only marginal
anisotropy in the inertial range and the statistics computed in
the different directions nearly coincided with each other except
for large scales $r/\eta > 300$. In figure \ref{fract_localslp} we
show the logarithmic derivative (the local slope) of the structure
functions for various fractional orders. If the local slope is
constant over a range of $r$, it provides the scaling exponent
$\zeta_{|n|}$. A rough estimate of the inertial range from
figure~\ref{fract_localslp} is $50 < r/\eta < 140$ and the error
estimate on the value scaling exponent is calculated as the
variance over this range. As expected, the smaller the absolute
order $|n|$, the more constant is the local slope, thus indicating
that our confidence level improves as the moment order decreases.
The scaling exponents and their uncertainty over the inertial
range are given for a range of fractional orders in
table~\ref{exponents}, column 6. The exponents calculated for all
three datasets display a good degree of agreement.

The scaling exponents computed in this way from the $1024^3$
simulation are plotted as a function of order $n$ in figure
\ref{fract_exps}. For comparison, the K41 exponents are also
shown. The measured exponents deviate from the K41 values for all
orders. This shows that the intermittency, until now thought to be
characteristic of only high-order moments, sampling `fat' tails of
the PDF of velocity increments, persists into the low-order
moments sampling the core of the PDF. Figure \ref{d-fract_exps}
shows the difference $(\zeta_{|n|} - n/3)$ from the K41 prediction
as a function of $n$. There is a linear approach to zero with a
slope of 0.056. Equivalently, the behavior near zero is
$\zeta_{|n|} = 0.38 n$ instead of $\zeta_{|n|} = n/3$.

Figure~\ref{relative_exps} shows the relative (percent) departure
of the measured scaling exponents from the self-similarity
prediction of K41, $(\zeta_{|n|} - n/3)/(n/3)$, is a smooth
function of $n$ in the range $-1 < n \leq 10$
(figure~\ref{relative_exps}). We have also presented the anomalous
exponents calculated from the multifractal $p$-model (Meneveau \&
Sreenivasan 1988) for comparison. The dependence on $n$ in the
range $-1 < n \leq 3$ is more or less linear and becomes weakly
quadratic for $n > 3$. The exact $4/5-$law result of K41 with
scaling exponent 1, takes the relative difference to zero at
$n=3$. A noteworthy point in this curve is at $n=0$. The relative
difference from K41 is not defined for $n=0$ but the experimental
and model values go smoothly through zero. We interpret this to
mean that anomalous scaling exists in the limit $n \rightarrow 0$
with relative departure from K41 of about $16\%$ (see the
$y$-intercept at $n=0$ in figure~\ref{relative_exps}).

\section{Discussion and conclusions}

The principal result of this paper is that the fractional
low-order moments have scaling exponents that are different from
$n/3$. In this sense, it is reasonable to consider that anomaly
exists in very low-order and negative moments. Thus, instead of
focusing entirely on high-order moments in search of an
explanation for intermittency, it may also be reasonable to
attempt to understand anomaly in low-order moments. This has the
advantage that rare events, whose universality may be in some
doubt, play far less of a role.

We should remark on a lingering uncertainty. To keep structure
functions real-valued for arbitrary orders, we can only consider,
in the range $-1 < n \le 2$, fractional moments of absolute valued
velocity increments defined through equation (\ref{Sn_a}). The
difference between the structure functions defined in (\ref{Sn})
and the corresponding ones defined for absolute-valued velocity
increments is by no means clear for large $|n|$. Our unpublished
work appears to indicate that the absolute-valued structure
functions have a larger scaling exponent than the classical ones
when $n$ is large and odd. It cannot therefore be dismissed
entirely that the departures from $n/3$ that one may observe for
small and fractional $n$ may merely suggest the possibility that
K41 does not somehow apply to absolute moments. Even if this is
our only conclusion, it is still new and thus of some interest.

\begin{acknowledgments}

\end{acknowledgments}

\end{document}